\documentclass[10pt]{article}
\usepackage[OE]{express}
\usepackage{textcomp}
\usepackage{pifont}
\begin{document}
\title{A 30 W, sub-kHz frequency-locked laser at 532 nm}

\author{Hao-Ze Chen,\authormark{1,2,3} Xiang-Pei Liu,\authormark{1,2,3} Xiao-Qiong Wang,\authormark{1,2} Yu-Ping Wu,\authormark{1,2} Yu-Xuan Wang,\authormark{1,2} Xing-Can Yao,\authormark{1,2,4} Yu-Ao Chen,\authormark{1,2} and Jian-Wei Pan\authormark{1,2}}

\address{
\authormark{1}Shanghai Branch, National Laboratory for Physical Sciences at Microscale and Department of Modern Physics, University of Science and Technology of China, Shanghai, 201315, China
\\
\authormark{2}CAS Center for Excellence and Synergetic Innovation Center in Quantum Information and Quantum Physics, University of Science and Technology of China, Hefei, Anhui 230026, China
\\
\authormark{3}These authors contributed equally to this work.\\
\authormark{4}Email: yaoxing@ustc.edu.cn\\
}

\begin{abstract}
We report on the realization of a high-power, ultranarrow-linewidth, and frequency-locked 532~nm laser system. The laser system consists of single-pass and intra-cavity second harmonic generation of a continuous-wave Ytterbium doped fiber laser at 1064~nm in the nonlinear crystal of periodically poled lithium niobate and lithium triborate, respectively. With 47~W infrared input, 30~W green laser is generated through the type I critical phase matching in the intracavity lithium triborate crystal. The laser linewidth is measured to be on the order of sub-kHz, which is achieved by simultaneously locking the single-pass frequency doubling output onto the iodine absorption line R69 (36-1) at 532~nm. Furthermore, the phase locking between the laser system and another slave 1064~nm laser is demonstrated with relative frequency tunability being up to 10~GHz. Our results completely satisfy the requirements of 532~nm laser for quantum simulation with ultracold atoms.
\end{abstract}

\section{Introduction}
Continuous-wave (cw) green laser sources play important roles in a broad range of scientific applications, such as gravitational wave detection\cite{gair2009probing}, optical frequency metrology\cite{ezekiel1968laser,eickhoff1995optical,levenson1972hyperfine} and fundamental physics\cite{burger1999dark,abo2001observation,yao2016observation}. With the rapid development of laser and nonlinear conversion techniques, high power green laser sources with dozens of watts have become commercially available. The state-of-the-art 532 nm laser system can produce up to 134~W of green laser light via intracavity second harmonic generation (SHG) with a conversion efficiency up to 90\%\cite{meier2010continuous}. This advanced technique benefits the next-generation gravitational-wave detectors such as the Evolved Laser Interferometer Space Antenna (eLISA)\cite{sesana2016prospects} and the Einstein Telescope (ET)\cite{punturo2010einstein}. On the other hand, single-frequency narrow-linewidth green lasers have been of great interest in optical frequency metrology for many years because the saturated absorption lines of iodine provide a set of relatively narrow and stable optical frequency references near 532~nm. Although optical lattice clocks have already brought the frequency accuracy down to the $10^{-19}$ level\cite{campbell2017fermi}, the iodine-stabilized green laser source is still extensively used because of its simplicity and robustness.

Moreover, the 532~nm laser are particularly suitable for quantum simulation with ultracold atoms\cite{bloch2012quantum} thanks to its short wavelength, high power, and availability; for instance, to realize a box trap\cite{gaunt2013bose,mukherjee2017homogeneous} or an optical lattice\cite{bakr2009quantum,simon2011quantum}. In the case of a box trap, the high power 532~nm laser is used to isolate the atom cloud in an enclosed space with zero light field intensity. It is essential to the study of homogenous ultracold quantum gases which shed a new light on the exploration of Kibble-Zurek theory\cite{kibble1976topology,zurek1985cosmological}, supersolid states\cite{bulgac2008unitary}, Bogoliubov theory of quantum depletion\cite{lopes2017quantum} and etc. In the case of an optical lattice, a laser with short wavelength has special advantage as it provides a short lattice spacing, which results in a fast tunneling rate characterized by lattice recoil frequency $v_{rec}=h/(8ma^2)$ ($a$ is the geometric lattice spacing, and $h$ is Planck's constant). This benefits the studies of the thermalization process and super-exchange dynamics\cite{trotzky2008time} of some heavy elements, such as Erbium\cite{baier2016extended} and Dysprosium\cite{lu2012quantum}. In addition, optical super-lattices can be realized by superimposing two standing waves created with 532~nm and 1064~nm lasers, paving the way for studying novel topological quantum matter\cite{lohse2017exploring}.

Although the green laser has shown great prospects in ultracold atoms experiments, there are still many obstacles in the way for driving these sophisticated experiments. Take optical lattices as an example, the phase noise and frequency drift of the laser could result in a random deformation of the potential structure, which induces dissipation among motional states\cite{ludlow2006systematic}. In addition, sufficient laser power results in a deep optical lattice which can prevent the atoms from tunneling away during long exposure of resonant imaging light\cite{bakr2009quantum,parsons2015site,cheuk2016observation}. Thus, an optical lattice benefits greatly if ultra-narrow linewidth, stable absolute frequency, and high power are simultaneously achieved in one green laser source. However, most available high power 532~nm laser sources are solid-state lasers with a typical linewidth of 5~MHz and wavelength tuning range of several GHz which do not meet the requirement.

In this work, we report on the development of a high-power, ultranarrow-linewidth, and frequency-locked 532~nm laser system. The main laser source is a cw high power fiber amplifier pumped by a Ytterbium doped single frequency DFB fiber laser with a maximum output power of 50~W at 1064~nm. The output of the fundamental laser is divided into two parts for frequency locking and generation of the high power 532~nm laser, respectively. To achieve the system's robustness and flexibility, the optical setups and servo loops for the two tasks are mutually independent. Frequency locking and linewidth narrowing are simultaneously demonstrated by implementing modulation transfer spectroscopy (MTS) to lock the laser onto one of the iodine absorption line R69(36-1) at 532~nm. The achieved frequency instability is about 30~kHz over a 1000~s measurement time with linewidth being on the order of sub-kHz, which are determined by the heterodyne beat-note measurements with two identical laser setups. High power 532~nm laser is yielded via intra-cavity SHG process. Although periodically poled crystals are quite efficient for SHG process, thermal lensing and thermal dephasing effect strongly affect the output power and long term stability. Here we choose Lithium triborate (LBO) crystal due to its large phase-matching bandwidth and damage threshold. Owing to its relatively low nonlinear coefficients, the enhanced cavity is carefully designed to achieve low astigmatism and high conversion efficiency. With a two stage locking scheme, we achieve more than 30~W green radiation with a long-term power stability better than 2.6\%. Furthermore, phase locking between the 532~nm laser and another slave 1064~nm laser is demonstrated, which is a key ingredient for realizing an optical superlattice.

\section{ Experimental implementation}

\begin{figure}[htbp]
  \centering
  \includegraphics[width=10cm]{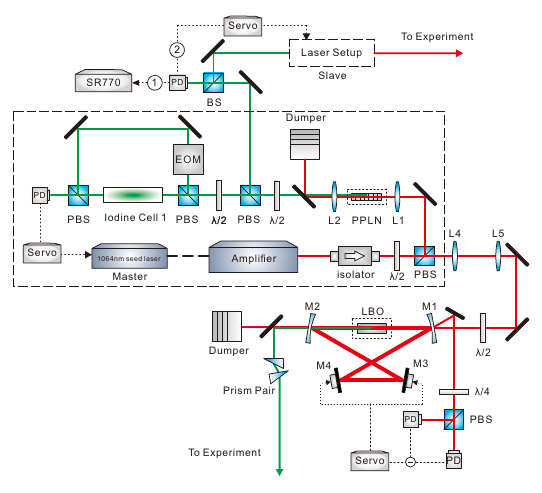}\\
  \caption{Experimental setup of the 532~nm laser system. The absolute frequency of the master laser at 1064 nm is stabilized by locking the laser frequency onto the transition line of molecular iodine. The main output of the infrared radiation is coupled into an external frequency doubling cavity to generate the high power green laser. The final green laser is delivered to vacuum chamber for atom trapping. The slave laser setup is used for verifying the locking performance by characterizing the beat-note, labelled as \ding{172}, and constructing the optical super-lattice, labelled as \ding{173}, respectively. The slave laser setup is identical with the master laser in the dotted box.}
  \label{f1}
\end{figure}

The overall layout of the experimental setup is shown in Fig.~\ref{f1}. About 8~mW seed laser (Koheras ADJUSTIK Y10) is injected into the fiber amplifier (YAR-50-1064-LP-SF), and more than 48~W of 1064~nm laser light is generated. The laser beam is divided into two by a half-wave plate (HWP) and a polarizing beam splitter (PBS) for frequency stabilization and high-power SHG generation, respectively. We first describe the experimental method for frequency stabilization. Using an aspheric lens L1, about 800~mW of 1064~nm laser is focused on a periodically poled lithium niobate (PPLN) crystal (Covesion, MSHG1064-1.0-20), which has dimensions of 20 mm$\times$1 mm$\times$10 mm. The crystal is heated to its quasi-phase matching temperature at 134~$^\circ$C with a stability of $\pm0.01~^\circ$C and about 25~mW 532~nm laser is generated. The green laser is collimated by another lens L2, passes through a dichroic mirror and a band-pass filter to remove the residual infrared laser component.

The resultant green laser is split into two parts for frequency locking and monitoring, respectively. To eliminate the Doppler background and residual amplitude modulation, the frequency stabilization is based on the MTS technique\cite{shirley1982modulation}. The probe beam and pump beam counter propagate through the iodine cell with an intensity ratio of 1:10. The pump beam is frequency modulated by a resonant Electro-Optic Modulator (EOM, Qubig, E0-Tx6M3-NIR), which is operated at 4.1~MHz. The iodine cell has Brewster-angled windows on the two ends and a length of 10~cm. To suppress the collisional broadening effect, the temperature of its cold-finger is stabilized at 5~$^\circ$C using a thermoelectric cooler (TEC) mounted on a copper heat sink. The iodine spectrum and the error signal for frequency locking are observed by the well-established Pound-Drever-Hall (PDH) method\cite{drever1983laser}. Experimentally, we find one of the transition R69 (36-1) of iodine (18792.53 $cm^{-1}$)\cite{gerstenkorn1978atlas} possesses the best spectral performance (slope and capture range), so it's adopted for frequency stabilization. Benefitting from the carefully chosen experimental parameters, the optimized slope of error signal near the transition center is up to 4.6~V per megahertz with the signal-to-noise ratio being more than 1000:1. Then, the error signal is sent into a PID controller (10~MHz bandwidth) and fed back to the PZT (20~kHz bandwidth) of the seed laser. We mention that the free running linewidth of the seed laser is about 3~kHz, and its phase noise is mainly distributed at low frequency. Thus, such a feedback system fulfills the requirement for the linewidth narrowing.

\begin{figure}[htbp]
  \centering
  \includegraphics[width=8cm]{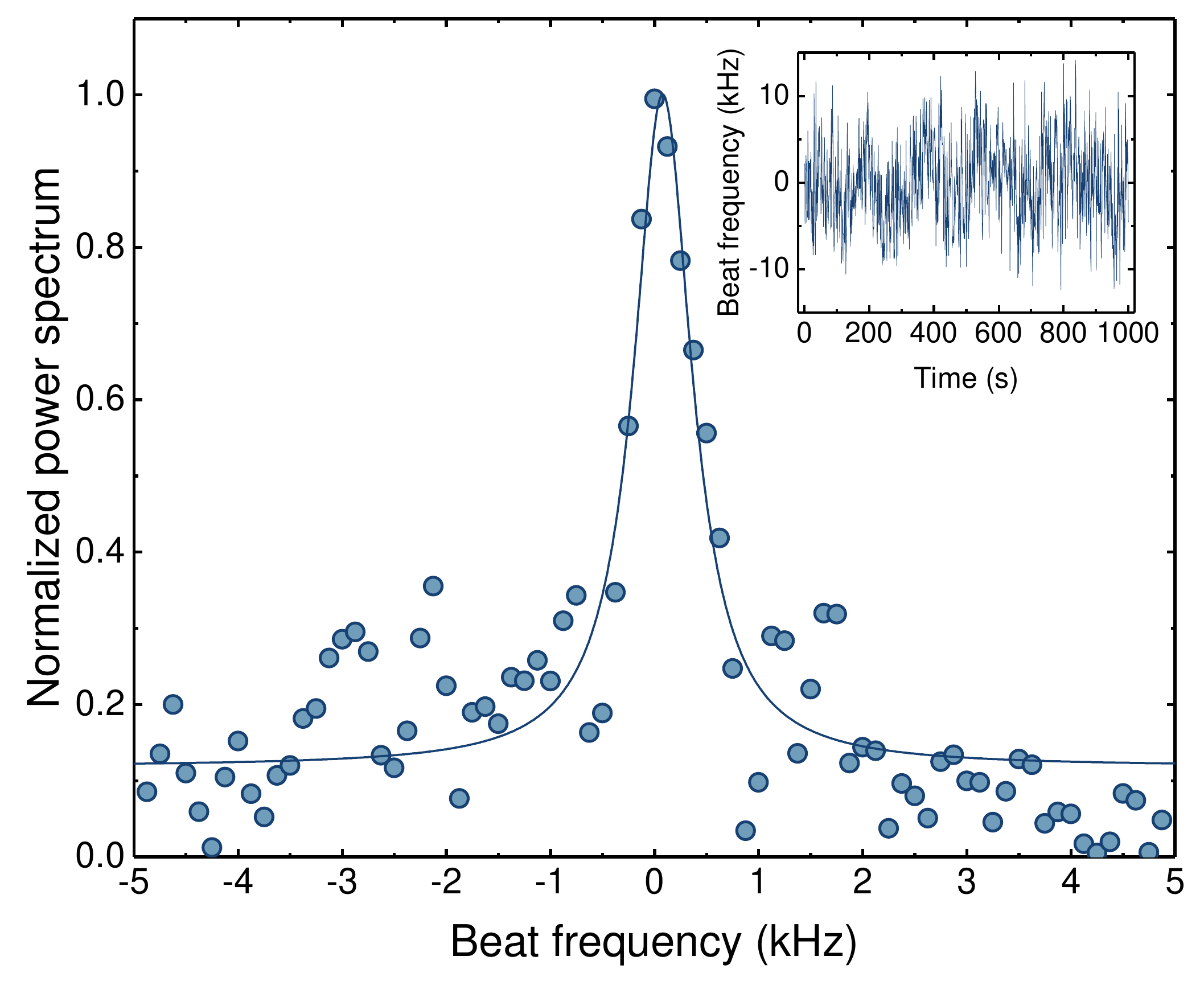}\\
  \caption{Power spectrum of optical beat-note signal between two identical SHG 532~nm lasers. The data is captured by a FFT analyser with 125~Hz resolution. The solid line indicates a Lorentz fit with FWHM of 679(73)~Hz. The central frequencies of the beat note signals are defined as zero for convenience. The inset shows a trace of the beat signal for a long-term of 1000~s.}
  \label{f2}
\end{figure}

To verify the locking performance of the laser linewidth and frequency stability, we build another independent 532~nm slave laser system. As shown in Fig.~\ref{f1},  for implementing the heterodyne beat-note linewidth measurements, the frequency stabilization setup of the slave laser is the same with the master laser. Both the master and slave lasers are locked onto exactly the same iodine molecular transition. By slightly tuning the zero point of the generated error signal, we obtain a non-zero beat-note frequency of less than 100~kHz, which could be recorded by a fast Fourier transform analyser. Figure~\ref{f2} gives a typical power spectrum density of the beat-note signal. A full-width at half-maximum (FWHM) of 679(73)~Hz is gained through Lorentz fitting. Next, to characterise the long-term stability of the laser frequency, we take a 1000~s record of the beat-note frequency with a counter gate time of 100~ms (see the inset of Fig.~\ref{f2}). The peak-to-peak frequency fluctuation is reduced to approximately 30~kHz. These results undoubtedly demonstrate that we successfully build a 532~nm laser system with a sub-kilohertz linewidth and high frequency stability.

Next, we present the experimental method for generating high power 532~nm laser.  About 47~W light power is coupled into an external frequency doubling cavity with a LBO crystal. As the thermal lensing effect has little influence on LBO crystal, we follow the Boyd and Kleinman's proposal\cite{boyd1968parametric} to design the cavity and maximize the SHG conversion efficiency. The bow-tie cavity is formed by two concave mirrors (M1: curvature radius 132~mm, M2: curvature radius 100~mm) and two flat mirrors (M3, M4). The concave mirrors are separated by 166~mm, while the flat mirrors are spaced by 144~mm. We position two parallel arms as close as possible (27~mm) to minimize the astigmatism; here, a folding angle of 10 degrees is achieved. The total cavity length is approximately 626~mm, resulting in a free spectral range of 480~MHz. The Type-I critical phase matching LBO crystal with dimensions of 3 mm$\times$3 mm$\times$20 mm is housed in a homemade copper holder and placed between two concave mirrors. The holder is operated at 25~$^\circ$C with a temperature stability of $\pm0.01~^\circ$C. The beam waist at the center of the crystal is approximately 27.5~$\mu$m. This gives a focusing parameter of $\xi=2.86$ which perfectly matches the optimum value of 2.84.

Due to the low effective nonlinearities $d_{eff}$ for the LBO crystal, mode matching of the SHG cavity, including the spatial and impedance matching\cite{kozlovsky1988efficient}, becomes crucial for the efficient generation of the high-power 532~nm laser. A set of spherical lenses (L4, L5) are used to match the eigenmode of the cavity. Moreover, an appropriate reflectance of input mirror M1 is selected to maximize the energy density of the fundamental laser. The optical power in the cavity $P_{cav}$ can be presented as

\begin{equation}
P_{cav}=\dfrac{T_1P_{in}}{(1-\sqrt{R_1(1-T_1R_2R_3R_4)(1-\eta P_{cav})})^2}
\end{equation}
where $P_{in}$ is the input power, $T_i$ and $R_i$ are the transmissivity and reflectivity of the mirrors, and $\eta$ is the single-pass frequency-doubling efficiency, respectively. From this equation, it's found that $R_1$ does not have a monotonic relationship with the cavity build-up factor $P_{cav}/P_{in}$. In the experiment, the reflectivity of the input mirror is determined to be 94\% for optimal impedance matching after computational simulation and several attempts.

\begin{figure}[htbp]
  \centering
  \includegraphics[width=13cm]{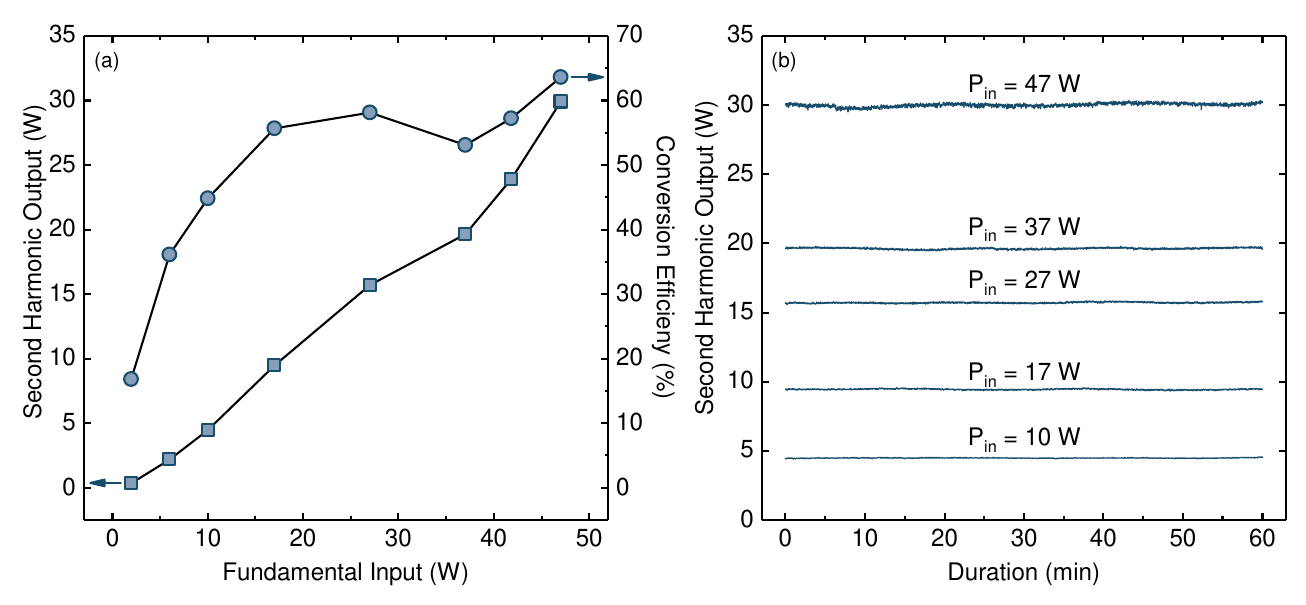}\\
  \caption{(a) SHG output power (black squares) and conversion efficiency (black circles) versus fundamental input power. The maximum output power is approximately 30 W, corresponding to a conversion efficiency of 64\%. (b) The long-term stability of SHG output with different fundamental input powers. The power variation is approximately 150 mW at 4.5 W output and increases to 780 mW at 30 W output.}
  \label{f3}
\end{figure}

Mode locking of the cavity is based on the H{\"a}nsch-Couillaud technique\cite{hansch1980laser} by detecting the phase difference between two polarization modes of the reflection light from the input mirror (M1). For achieving both fast feedback and wide capture range of the cavity length, the plain mirrors (M3, M4) are mounted on two piezo actuators with different stokes and response bandwidth, respectively. The generated error signal is sent into a PID controller (Toptica, FALC110), which slow and fast outputs are being fed back to the corresponding piezo actuators, respectively. At the output end of the cavity, a dichroic mirror is employed to separate the generated 532~nm laser from the residual 1064~nm radiation. Figure~\ref{f3}(a) shows the 532~nm laser power and SHG conversion efficiency as a function of the fundamental infrared laser power. For each input power level, we optimize the cavity coupling as well as the electronic control systems. The SHG output scales almost linearly with the fundamental input power while the conversion efficiency starts to saturate at around 60\%. At the maximum input of 47~W, 30~W of 532~nm laser is obtained, corresponding to a conversion efficiency of 64\%. The limited fundamental power precludes us from achieving higher second harmonic output. We also measure the beam quality of the 532~nm laser with a CCD camera mounted on a motorized precision translation stage. The obtained M$^2$ value is 1.12 (1.04) along the x (y) axis, respectively.

To investigate the long-term stability of the SHG output, the 532~nm laser power is measured over 1 hour with different infrared input powers (see Fig.~\ref{f3}(b)). The results show a good power stability of our 532~nm laser with the peak-to-peak fluctuation being 150~mW at 4.5~W output and 780~mW at 30 W output, corresponding to a power stability of 2.6\%. Furthermore, we observe a long-term periodical fluctuation on a time-scale of about 23 minutes in each power scale, mainly due to the power fluctuation of the fundamental input.  The high power and ultranarrow linewith of the laser system allows us to build up a deep 532 nm optical lattice with ultralow noise. For instance, the atom heating effect that induced by the laser phase noise can be greatly suppressed by three orders of magnitude compared with those commercial laser sources.

\begin{figure}[htbp]
  \centering
  \includegraphics[width=13cm]{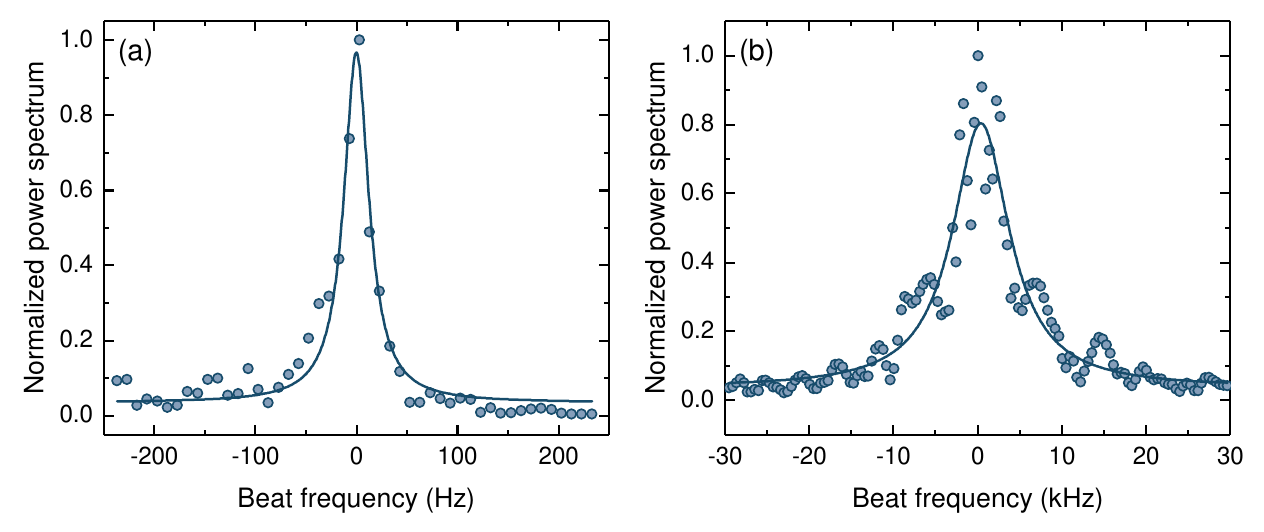}\\
  \caption{Power spectral density profile of beat-note signal in the phase locking (a) and free running case (b). The solid lines indicate Lorentz fits with FWHM of 30(2)~Hz (a) and 8.2(5)~kHz, respectively. The central frequencies of the beat note signals are defined as zero for convenience.}
  \label{f4}
 \end{figure}

Finally, we show that the achieved 532~nm laser can be phase locked with another fiber laser at 1064~nm with tunable frequency difference, which is a key ingredient for constructing an optical super-lattice. In this case, the slave laser is no longer locked onto the iodine spectrum, but phase locked to the master seed laser by an offset phase lock servo. The rapid and precise control of the relative frequency between the two laser is achieved with a 10 GHz tunability using a homemade PC-controlled DDS.  Figure~\ref{f4}(a) shows the measurement result of the beat-note signal after phase locking. The power spectral density profile of the beat note signal is fitted with Lorentzian function, giving a FWHW of 30(2)~Hz. Since the frequency fluctuations cause the random change of the lattice potential, the stability of frequency difference between the two lasers is very crucial for a superlattice experiment. Therefore, we also measure the long-term stability of the frequency difference and the peak-to-peak fuctuation is 8~kHz over 1000~s. Considering a typical length of 30~cm between the retro-reflecting mirror and the atom cloud, the phase fluctuation of the superlattice can be suppressed to $3.2\times10^{-5}\pi$, perfectly meeting the experimental requirements. For comparison, the power spectrum density of the free-running case is also given in Fig.~\ref{f4}(b), where a FWHW of 8.2(5)~kHz is obtained with dozens of MHz frequency fluctuations.

 Although these results satisfy the experimental requirements on the 532~nm laser, the system still holds great potential for improving. First, we can achieve narrower linewidth and less frequency drift of the 532~nm laser with a better optical frequency standard. The typical linewidth of the iodine spectrum near 532~nm is approximately 200~kHz\cite{cheng2002sub}. However, the pressure and power broadening effects prevent accurate determination of the center of the transition line. Thus, the temperature of the cold finger and optical power intensity should be further reduced to suppress these effects. Second, additional locking loops can be set up for frequency and intensity stabilization, respectively. For example, we can insert an AOM before the green laser enter the path of frequency locking, the output signal of a fast PID controller can be fed back to the AOM's driving frequency, which has much higher frequency response bandwidth\cite{jiang2011making}. In addition, the power stability and relative intensity noise of the 532~nm laser is expected to be around 1\textperthousand~and less than -120~dBc/Hz by dynamically controlling the AOM's modulation depth\cite{blatt2015low}.

\section{Conclusion}
In summary, we combine single-pass and intra-cavity SHG of a cw Ytterbium doped fiber laser to produce up to 30~W 532~nm laser with a linewidth at sub-kHz level. The ultra-narrow linewidth and absolute frequency stability of the laser is achived by performing modulation transfer spectroscopy of molecular iodine and verified through the heterdyne beating measurements. Using a special designed SHG cavity with LBO crystal, the high-power green laser is generated with the overall conversion efficiency being about 64\%.  By selecting appropriate cavity configurations, we reduce the influence of the astigmatism of optical mode and achieve a beam quality factor M$^2$ of 1.12 (1.04) along the x (y) axis, respectively. Future work will improve further upon the reported power stability, intensity noise, laser linewidth, and frequency drift. Our work provides an ideal 532~nm laser system for the quantum simulation with ultracold atoms. With its high power, frequency stability and tunability, we expect a better study of the Bose and Fermi superfluid mixture~\cite{yao2016observation} in the 532~nm optical lattices.

\section*{Funding}
National Natural Science Foundation of China (NSFC) (11874340, 11425417); National Key R\&D Program of China (2018YFA0306501); Chinese Academy of Sciences (CAS); Anhui Initiative in Quantum Information Technologies; Fundamental Research Funds for the Central Universities (WK2340000081).

\bibliographystyle{osajnl}
\bibliography{532laser}

\begin{thebibliography}{10}
\newcommand{\enquote}[1]{``#1''}

\bibitem{gair2009probing}
J.~R. Gair, I.~Mandel, A.~Sesana, and A.~Vecchio, \enquote{Probing seed black
  holes using future gravitational-wave detectors,} Class. Quantum Grav.
  \textbf{26}, 204009 (2009).

\bibitem{ezekiel1968laser}
S.~Ezekiel and R.~Weiss, \enquote{Laser-induced fluorescence in a molecular
  beam of iodine,} Phys. Rev. Lett. \textbf{20}, 91--93 (1968).

\bibitem{eickhoff1995optical}
M.~Eickhoff and J.~Hall, \enquote{Optical frequency standard at 532 nm,} {IEEE}
  Trans. Instrum. Meas. \textbf{44}, 155--158 (1995).

\bibitem{levenson1972hyperfine}
M.~D. Levenson and A.~L. Schawlow, \enquote{Hyperfine interactions in molecular
  iodine,} Phys. Rev. A \textbf{6}, 10--20 (1972).

\bibitem{burger1999dark}
S.~Burger, K.~Bongs, S.~Dettmer, W.~Ertmer, K.~Sengstock, A.~Sanpera, G.~V.
  Shlyapnikov, and M.~Lewenstein, \enquote{Dark solitons in {Bose-Einstein}
  condensates,} Phys. Rev. Lett. \textbf{83}, 5198--5201 (1999).

\bibitem{abo2001observation}
J.~R. Abo-Shaeer, C.~Raman, J.~M. Vogels, and W.~Ketterle, \enquote{Observation
  of vortex lattices in {Bose-Einstein} condensates,} Science \textbf{292},
  476--479 (2001).

\bibitem{yao2016observation}
X.-C. Yao, H.-Z. Chen, Y.-P. Wu, X.-P. Liu, X.-Q. Wang, X.~Jiang, Y.~Deng,
  Y.-A. Chen, and J.-W. Pan, \enquote{Observation of coupled vortex lattices in
  a mass-imbalance {Bose} and {Fermi} superfluid mixture,} Phys. Rev. Lett.
  \textbf{117}, 145301 (2016).

\bibitem{meier2010continuous}
T.~Meier, B.~Willke, and K.~Danzmann, \enquote{Continuous-wave single-frequency
  532 nm laser source emitting 130 {W} into the fundamental transversal mode,}
  Opt. Lett. \textbf{35}, 3742 (2010).

\bibitem{sesana2016prospects}
A.~Sesana, \enquote{Prospects for multiband gravitational-wave astronomy after
  {GW}150914,} Phys. Rev. Lett. \textbf{116}, 231102 (2016).

\bibitem{punturo2010einstein}
M.~Punturo, M.~Abernathy, F.~Acernese, B.~Allen, N.~Andersson, K.~Arun,
  F.~Barone, B.~Barr, M.~Barsuglia, M.~Beker, N.~Beveridge, S.~Birindelli,
  S.~Bose, L.~Bosi, S.~Braccini, C.~Bradaschia, T.~Bulik, E.~Calloni, G.~Cella,
  E.~C. Mottin, S.~Chelkowski, A.~Chincarini, J.~Clark, E.~Coccia, C.~Colacino,
  J.~Colas, A.~Cumming, L.~Cunningham, E.~Cuoco, S.~Danilishin, K.~Danzmann,
  G.~D. Luca, R.~D. Salvo, T.~Dent, R.~D. Rosa, L.~D. Fiore, A.~D. Virgilio,
  M.~Doets, V.~Fafone, P.~Falferi, R.~Flaminio, J.~Franc, F.~Frasconi,
  A.~Freise, P.~Fulda, J.~Gair, G.~Gemme, A.~Gennai, A.~Giazotto,
  K.~Glampedakis, M.~Granata, H.~Grote, G.~Guidi, G.~Hammond, M.~Hannam,
  J.~Harms, D.~Heinert, M.~Hendry, I.~Heng, E.~Hennes, S.~Hild, J.~Hough,
  S.~Husa, S.~Huttner, G.~Jones, F.~Khalili, K.~Kokeyama, K.~Kokkotas,
  B.~Krishnan, M.~Lorenzini, H.~L\"{u}ck, E.~Majorana, I.~Mandel, V.~Mandic,
  I.~Martin, C.~Michel, Y.~Minenkov, N.~Morgado, S.~Mosca, B.~Mours,
  H.~M\"{u}ller{\textendash}Ebhardt, P.~Murray, R.~Nawrodt, J.~Nelson,
  R.~Oshaughnessy, C.~D. Ott, C.~Palomba, A.~Paoli, G.~Parguez, A.~Pasqualetti,
  R.~Passaquieti, D.~Passuello, L.~Pinard, R.~Poggiani, P.~Popolizio, M.~Prato,
  P.~Puppo, D.~Rabeling, P.~Rapagnani, J.~Read, T.~Regimbau, H.~Rehbein,
  S.~Reid, L.~Rezzolla, F.~Ricci, F.~Richard, A.~Rocchi, S.~Rowan,
  A.~R\"{u}diger, B.~Sassolas, B.~Sathyaprakash, R.~Schnabel, C.~Schwarz,
  P.~Seidel, A.~Sintes, K.~Somiya, F.~Speirits, K.~Strain, S.~Strigin,
  P.~Sutton, S.~Tarabrin, A.~Th\"{u}ring, J.~van~den Brand, C.~van Leewen,
  M.~van Veggel, C.~van~den Broeck, A.~Vecchio, J.~Veitch, F.~Vetrano,
  A.~Vicere, S.~Vyatchanin, B.~Willke, G.~Woan, P.~Wolfango, and K.~Yamamoto,
  \enquote{The {Einstein} telescope: a third-generation gravitational wave
  observatory,} Class. Quantum Grav. \textbf{27}, 194002 (2010).

\bibitem{campbell2017fermi}
S.~L. Campbell, R.~B. Hutson, G.~E. Marti, A.~Goban, N.~D. Oppong, R.~L.
  McNally, L.~Sonderhouse, J.~M. Robinson, W.~Zhang, B.~J. Bloom, and J.~Ye,
  \enquote{A {Fermi}-degenerate three-dimensional optical lattice clock,}
  Science \textbf{358}, 90--94 (2017).

\bibitem{bloch2012quantum}
I.~Bloch, J.~Dalibard, and S.~Nascimb{\`{e}}ne, \enquote{Quantum simulations
  with ultracold quantum gases,} Nat. Phys. \textbf{8}, 267--276 (2012).

\bibitem{gaunt2013bose}
A.~L. Gaunt, T.~F. Schmidutz, I.~Gotlibovych, R.~P. Smith, and Z.~Hadzibabic,
  \enquote{{Bose-Einstein} condensation of atoms in a uniform potential,} Phys.
  Rev. Lett. \textbf{110}, 200406 (2013).

\bibitem{mukherjee2017homogeneous}
B.~Mukherjee, Z.~Yan, P.~B. Patel, Z.~Hadzibabic, T.~Yefsah, J.~Struck, and
  M.~W. Zwierlein, \enquote{Homogeneous atomic {Fermi} gases,} Phys. Rev. Lett.
  \textbf{118}, 123401 (2017).

\bibitem{bakr2009quantum}
W.~S. Bakr, J.~I. Gillen, A.~Peng, S.~F\"{o}lling, and M.~Greiner, \enquote{A
  quantum gas microscope for detecting single atoms in a {Hubbard}-regime
  optical lattice,} Nature \textbf{462}, 74--77 (2009).

\bibitem{simon2011quantum}
J.~Simon, W.~S. Bakr, R.~Ma, M.~E. Tai, P.~M. Preiss, and M.~Greiner,
  \enquote{Quantum simulation of antiferromagnetic spin chains in an optical
  lattice,} Nature \textbf{472}, 307--312 (2011).

\bibitem{kibble1976topology}
T.~W.~B. Kibble, \enquote{Topology of cosmic domains and strings,} J. Phys. A:
  Math. Gen. \textbf{9}, 1387--1398 (1976).

\bibitem{zurek1985cosmological}
W.~H. Zurek, \enquote{Cosmological experiments in superfluid helium?} Nature
  \textbf{317}, 505--508 (1985).

\bibitem{bulgac2008unitary}
A.~Bulgac and M.~M. Forbes, \enquote{Unitary {Fermi} supersolid: The
  {Larkin-Ovchinnikov} phase,} Phys. Rev. Lett. \textbf{101}, 215301 (2008).

\bibitem{lopes2017quantum}
R.~Lopes, C.~Eigen, N.~Navon, D.~Cl{\'{e}}ment, R.~P. Smith, and Z.~Hadzibabic,
  \enquote{Quantum depletion of a homogeneous {Bose-Einstein} condensate,}
  Phys. Rev. Lett. \textbf{119}, 190404 (2017).

\bibitem{trotzky2008time}
S.~Trotzky, P.~Cheinet, S.~F\"{o}lling, M.~Feld, U.~Schnorrberger, A.~M. Rey,
  A.~Polkovnikov, E.~A. Demler, M.~D. Lukin, and I.~Bloch,
  \enquote{Time-resolved observation and control of superexchange interactions
  with ultracold atoms in optical lattices,} Science \textbf{319}, 295--299
  (2008).

\bibitem{baier2016extended}
S.~Baier, M.~J. Mark, D.~Petter, K.~Aikawa, L.~Chomaz, Z.~Cai, M.~Baranov,
  P.~Zoller, and F.~Ferlaino, \enquote{Extended {Bose-Hubbard} models with
  ultracold magnetic atoms,} Science \textbf{352}, 201--205 (2016).

\bibitem{lu2012quantum}
M.~Lu, N.~Q. Burdick, and B.~L. Lev, \enquote{Quantum degenerate dipolar
  {Fermi} gas,} Phys. Rev. Lett. \textbf{108}, 215301 (2012).

\bibitem{lohse2017exploring}
M.~Lohse, C.~Schweizer, H.~M. Price, O.~Zilberberg, and I.~Bloch,
  \enquote{Exploring {4D} quantum hall physics with a {2D} topological charge
  pump,} Nature \textbf{553}, 55--58 (2018).

\bibitem{ludlow2006systematic}
A.~D. Ludlow, M.~M. Boyd, T.~Zelevinsky, S.~M. Foreman, S.~Blatt, M.~Notcutt,
  T.~Ido, and J.~Ye, \enquote{Systematic study of the $^{87}${Sr} clock
  transition in an optical lattice,} Phys. Rev. Lett. \textbf{96}, 033003
  (2006).

\bibitem{parsons2015site}
M.~F. Parsons, F.~Huber, A.~Mazurenko, C.~S. Chiu, W.~Setiawan,
  K.~Wooley-Brown, S.~Blatt, and M.~Greiner, \enquote{Site-resolved imaging of
  {Fermionic} $^6${Li} in an optical lattice,} Phys. Rev. Lett. \textbf{114},
  213002 (2015).

\bibitem{cheuk2016observation}
L.~W. Cheuk, M.~A. Nichols, K.~R. Lawrence, M.~Okan, H.~Zhang, and M.~W.
  Zwierlein, \enquote{Observation of {2D Fermionic Mott} insulators of
  $^{40}${K} with single-site resolution,} Phys. Rev. Lett. \textbf{116},
  235301 (2016).

\bibitem{shirley1982modulation}
J.~H. Shirley, \enquote{Modulation transfer processes in optical heterodyne
  saturation spectroscopy,} Opt. Lett. \textbf{7}, 537 (1982).

\bibitem{drever1983laser}
R.~W.~P. Drever, J.~L. Hall, F.~V. Kowalski, J.~Hough, G.~M. Ford, A.~J.
  Munley, and H.~Ward, \enquote{Laser phase and frequency stabilization using
  an optical resonator,} Appl. Phys. B \textbf{31}, 97--105 (1983).

\bibitem{gerstenkorn1978atlas}
S.~Gerstenkorn and P.~Luc, \enquote{Atlas du spectre d'absorption de la
  molecule d'iode 14800-20000 cm$^{-1}$,} {Paris: Editions du Centre National
  de la Recherche Scientifique (CNRS)}, 1978  (1978).

\bibitem{boyd1968parametric}
G.~D. Boyd and D.~A. Kleinman, \enquote{Parametric interaction of focused
  gaussian light beams,} J. Appl. Phys. \textbf{39}, 3597--3639 (1968).

\bibitem{kozlovsky1988efficient}
W.~Kozlovsky, C.~Nabors, and R.~Byer, \enquote{Efficient second harmonic
  generation of a diode-laser-pumped {CW Nd:YAG} laser using monolithic
  {MgO:LiNbO}/sub 3/ external resonant cavities,} {IEEE} J. Quantum Electron.
  \textbf{24}, 913--919 (1988).

\bibitem{hansch1980laser}
T.~Hansch and B.~Couillaud, \enquote{Laser frequency stabilization by
  polarization spectroscopy of a reflecting reference cavity,} Opt. Commun.
  \textbf{35}, 441--444 (1980).

\bibitem{cheng2002sub}
W.-Y. Cheng, L.~Chen, T.~H. Yoon, J.~L. Hall, and J.~Ye, \enquote{Sub-doppler
  molecular-iodine transitions near the dissociation limit (523{\textendash}498
  nm),} Opt. Lett. \textbf{27}, 571 (2002).

\bibitem{jiang2011making}
Y.~Y. Jiang, A.~D. Ludlow, N.~D. Lemke, R.~W. Fox, J.~A. Sherman, L.-S. Ma, and
  C.~W. Oates, \enquote{Making optical atomic clocks more stable with
  $10^{-16}$-level laser stabilization,} Nat. Photonics \textbf{5}, 158--161
  (2011).

\bibitem{blatt2015low}
S.~Blatt, A.~Mazurenko, M.~F. Parsons, C.~S. Chiu, F.~Huber, and M.~Greiner,
  \enquote{Low-noise optical lattices for ultracold $^6${Li},} Phys. Rev. A
  \textbf{92}, 021402 (2015).

\end{thebibliography}

\end{document}